\documentclass[preprin,showpacs,preprintnumbers,eqsecnum,amsmath,amssymb,nofootinbib]{revtex4}
\usepackage{graphics,epsfig,xcolor}
\usepackage{graphicx}
\usepackage [latin1]{inputenc}

\begin{document}
\baselineskip=0.8 cm
\title{Acoustic Schwarzschild-like black hole and analogous Hawking radiation in Einstein-bumblebee gravity}
\author{Chikun Ding$^{1,2}$}\thanks{Email: dingchikun@163.com}
\author{Yuebing Zhou$^{1}$}\author{Yu Shi$^{1}$}
\author{Yuehua Xiao$^{1}$}\author{Jiliang Jing$^{2}$}
\affiliation{$^1$Department of Physics, Huaihua University, Huaihua, 418008, P. R. China\\
$^2$Key Laboratory of Low Dimensional Quantum Structures and Quantum Control of Ministry of Education,
 Hunan Normal University, Changsha, Hunan 410081, P. R. China}
\vspace*{0.2cm}
\begin{abstract}
\baselineskip=0.6 cm

\begin{center}{\bf Abstract}\end{center}
We derive an acoustic Schwarzschild-like black hole and study its analogous Hawking radiation in Einstein-bumblebee theory by using Gross-Pitaevskii theory in the background of Schwarzschild-like spacetime. This black hole have two kinds of event horizons: one is the gravitational horizon $r_h=2M$(black hole horizon), the others are the internal and external acoustic horizons $2M<r_{ac-}\leq4M,\;r_{ac}\geq4M$(dumb hole horizon) which depend on the tuning parameter $\xi$ and Lorentz violation coupling constant $\ell$. The corresponding Hawking temperature $T_h$ depends only on $\ell$, however $T_{ac}$ depends on $\ell$ and $\xi$. The dumb hole temperature $T_{ac}$ has a maximum $T_{acm}=T_h/8$. When considering the scalar particles radiating from the dumb hole, we find that in low frequency, the grey-body factor increases with the tuning parameter $\xi$, but decreases with $\ell$. In the high frequency, it approaches to $1$. Its energy emission rate is similar to the Planck's black spectrum where there has a maximum at a given peak frequency $\omega_m$.  The energy emission rate also increases with the tuning parameter $\xi$, but decreases with $\ell$. This opposite dependence relationship is the most experimentally valuable feature and  can directly benchmark against observations of real astrophysical black holes.
\end{abstract}

 \maketitle
\vspace*{0.2cm}
\section{Introduction}
As the most extreme objects in general relativity, black holes  have been the most profound testing grounds for fundamental physics, bridging the classical description of curved spacetime and the yet incomplete framework of quantum gravity. In 1974, Hawking \cite{hawking,hawking248} predicted that a Schwarzschild black hole can spontaneously emit thermal radiation from its event horizon. After that, a lot of theoretical studies of Hawking radiation have reshaped our understanding of black hole thermodynamics, quantum field theory and the deep connection between gravity and quantum mechanics. Although several decades passed, the experimental verification of Hawking radiation is still elusive due to that the characteristic Hawking temperature of an astrophysical stellar-mass black hole is far lower than the temperature of the cosmic microwave background. This long-standing observational barrier has driven physicists to develop analogue gravity systems, which reproduce the geometric structure of black hole horizons in laboratory-accessible physical media, to test the core predictions of Hawking radiation in a controlled experimental setting.

In 1981, W. G. Unruh  \cite{unruh} first  proposed an analogue gravity systems---an acoustic black hole can be constructed by driving a fluid to form a supersonic transonic flow, where the fluid velocity exceeds the local speed of sound at a well-defined interface---acoustic horizon. It shares an identical mathematical structure with the event horizon of a general relativistic black hole: sound waves (phonons) propagating against the fluid flow cannot escape the supersonic region, just as no massive or massless particle can escape the gravitational horizon of a real black hole. The features including horizon, ergosphere and Hawking radiation of these acoustic black holes were explored in \cite{visser}, inspired by which more efforts on the analogue Hawking radiation were made in \cite{zhang2011,vieira,guo}. The recent experimental realizations of sonic black hole were conducted in a Bose-Einstein condensate \cite{lahav,lahav2010,steinhauer}, and the thermal Hawking radiation and the corresponding temperature in an analogue black hole were observed in Refs \cite{nova,isoard,steinhauer2016}. Being able to mimic kinematics aspects of general relativity and realize astrophysical phenomena in the laboratory, the analogue models of gravity were proposed to explore the quantum nature of gravity.

Though the standard acoustic black hole is built on the assumption of exact Lorentz invariance, modern quantum gravity frameworks and condensed matter theory both strongly suggest that Lorentz symmetry cannot be an exact fundamental symmetry at all energy scales \cite{mattingly,camelia,kostelecky}. At energies approaching the Planck scale, a wide class of effective field theories, such as the standard-model extension \cite{colladay,colladay1998,kostelecky}, predict small and measurable deviations from exact Lorentz invariance, which modify the dispersion relation of quantum fields, reshape the near-horizon geometric structure, and introduce non-trivial corrections to Hawking temperature, quasinormal modes, greybody factors and the full radiation spectrum. One of the most physically motivated and mathematically tractable frameworks for describing such effects is the bumblebee gravity model, first studied by Kosteleck\'{y} and Dickinson \cite{kostelecky198940,kostelecky39,dickinson}. In this model, the Lagrange density still preserves Lorentz invariance,  but the local Lorentz symmetry is just spontaneously broken by a potential $V(B^\mu B_\mu\mp b^2)$, where $B^\mu$ is a dynamic vector field with a non-zero vacuum expectation value $b^\mu$, without introducing pathologies such as ghost modes or causality violation. Unlike many {\it ad hoc} Lorentz-violating models, bumblebee gravity preserves diffeomorphism invariance, naturally generates Schwarzschild-like black hole solution, and has well-defined observational constraints from solar system tests \cite{casana}.

The first work on acoustic black hole with Lorentz symmetry breaking had been done by Visser \cite{visser}, in which Lorentz invariance was broken by the viscosity of a given fluid. Recently, Anacleto {\it et al} construct some modified acoustic black holes by considering the Abelian Higgs model in Lorentz-violating \cite{anacleto2023,anacleto2011,campos} and noncommutative \cite{anacleto2012} background in flat Minkowski spacetime.
However, gravity is not only able be mimicked in flat spacetime, but also in curved spacetime.  Because the real black holes in our galaxies  could be in the bath of  the cosmological microwave background. If this real black hole is Schwarzschild, then at the location $r_s=6GM$, the escape velocity $v_s=c/\sqrt{3}$, where $c$ is the speed of light in vacuum. In the region $r<r_s$, relativistic sound waves of thermal radiation cannot escape \cite{ge2019,hartle}. In addition to the cosmological microwave, transonic accretion disks and the condensation of some quantum fluids around the real black hole also provide a scenario realizing this analogue gravity \cite{moncrief,das,abraham}.

In 2019, Ge {\it et al} \cite{ge2019} constructed an acoustic black hole in general curved spacetime via the relativistic Gross-Pitaevskii theory. In this paper, we for the first time consider construction of an acoustic black hole in a curved spacetime with Lorentz invariance violation and study its Hawking radiation.
The rest of the paper is organized as follows.   In Sec. II, we construct an acoustic black hole in Schwarzschild-like black hole spacetime via Gross-Pitaevskii theory. In Sec. III, we study the greybody factor and the energy emission rate of its Hawking radiation to obtain the effect of Lorentz invariant violation on the phonon radiation. Sec. IV gives a summary and discussion. Appendix gives a canonical acoustic black hole in curved spacetime with Lorentz symmetry breaking.

\section{Construction of an acoustic black hole in Einstein-bumblebee gravity theory}

In this section, we construct an acoustic black hole solution in the background of a curved spacetime---Schwarzschild-like black hole spacetime with the Lorentz symmetry breaking orientation.

In the Einstein-bumblebee gravity theory, Casana {\it et al} \cite{casana} found an
exact Schwarzschild-like black hole solution which reads,
\begin{eqnarray}\label{metricbum}
&&ds_{bg}^2=g_{\mu\nu}dx^\mu dx^\nu=-f(r)dt^2+\frac{1+\ell}{f(r)}dr^2+r^2d\Omega^2,\quad f(r)=1-\frac{2M}{r},
\end{eqnarray}
where $M$ is the black hole mass and $\ell$ is the Lorentz violation coupling constant. In the limit of $M\rightarrow0$, the Kretschmann invariant doesn't vanish, $R_{\mu\nu\rho\tau}R^{\mu\nu\rho\tau}=4\ell^2/(1+\ell)^2r^4$, which indicates that the Lorentz violation effect deforms the given spacetime.
We use it as the static spherically symmetric  background spacetime to construct an acoustic black hole from field theory.

We consider a superfluid described by Gross-Pitaevskii equation in the Schwarzschild-like black hole spacetime and assume that the matter field has no back reaction to the background spacetime in the probe limit.   The relativistic Gross-Pitaevskii theory yields the action \cite{gross,pitaevskii,fagnocchi}
\begin{eqnarray}\label{}
&&S=\int d^4x\sqrt{-g}\left(|\partial_\mu\varphi|^2+m^2|\varphi|^2-\frac{b}{2}|\varphi|^4\right),
\end{eqnarray}
where $\sqrt{-g}$ is the determinant of the Schwarzschild-like black hole metric $g_{\mu\nu}$, the complex scalar order parameter $\varphi$ is  a macroscopic wave function  describing the quantum condensate. $m^2$ is a temperature-dependent mass square parameter assumed as $m^2\sim(T-T_c)$, where $T_c$ is the critical temperature of the superfluid. $b$ is a  self-interaction coupling constant and will be related to the sound wave speed. The equation of motion for $\varphi$ is
\begin{eqnarray}\label{}
&&\square\varphi+m^2\varphi-b|\varphi|^2\varphi=0.
\end{eqnarray}
This relativistic Gross-Pitaevskii equation is the nonlinear wave equation satisfied by the macroscopic wave function that describes a Bose-Einstein condensate. It also can  be  seen as the nonlinear extension of Klein-Gordon equation.
By using Madelung representation of the condensation $\varphi=\sqrt{\rho}e^{i\theta}$, the fluid background solution can be assumed as $(\rho_0,\theta_0).$ A dumb hole can be obtained by perturbations around this background:
\begin{eqnarray}\label{}
&&\rho=\rho_0+\epsilon\rho_1,\quad \theta=\theta_0+\epsilon\theta_1,
\end{eqnarray}
where $\epsilon$ is a small constant. In the long-wavelength limit, the quantum potential terms can be neglected and two equations---the leading order and the weak excitations can be extracted which govern the propagation of phase fluctuation in a homogeneous stationary condensate:
\begin{eqnarray}\label{rhom}
&&b\rho_0=m^2-v_\mu v^\mu,\quad \frac{1}{\sqrt{-\mathcal{G}}}\partial_\mu(\sqrt{-\mathcal{G}}\mathcal{G}^{\mu\nu}\partial_\nu\theta_1)=0,
\end{eqnarray}
where $v_\mu$ are the four-velocity of the fluid, $v_t=-\partial_t\theta_0, v^t=g^{tt}v_t, v_i=\partial_i\theta_0, v^i=g^{ii}v_i$, and $\mathcal{G}_{\mu\nu}$ is the metric of the new acoustic black hole,
\begin{eqnarray}\label{dtdr}
&&ds^2=\mathcal{G}_{\mu\nu}dx^\mu dx^\nu=H\big[g_{tt}(c_s^2-v^rv_r)dt^2-2v_tv_rdtdr+(c_s^2-v_tv^t)g_{rr}dr^2+(c_s^2-v_\mu v^\mu)r^2d\Omega^2\big],
\end{eqnarray}
with $H=c_s/\sqrt{c_s^2-v_\mu v^\mu}, c_s^2=b\rho_0/2$, and $v_\theta=v_\varphi=0$. If one applies flat background spacetime, i.e., sets that $g_{tt}=-1,\;g_{ii}=1$ and $c_s\rightarrow c_s/v_t, v_i\rightarrow v_i/v_t$, it becomes that
\begin{eqnarray}\label{metricrel}
&&ds^2=\tilde H\big[-(c_s^2-v^2_r)dt^2-2v_rdtdr+(c_s^2+1)dr^2+(1+c_s^2-v_r^2)r^2d\Omega^2\big],
\end{eqnarray}
with $\tilde H=b\rho_0/2c_s\sqrt{1+c_s^2-v_r^2}$. It is easy to see that this metric is the same as that of the acoustic black hole for relativistic fluids, see Eq. (2.17) in Ref. \cite{ge2010} when $dr^2+r^2d\Omega^2\rightarrow d\vec{x}^2$ and $v_1=v_r, v_2=v_3=0$.
In the non-relativistic limit, $c_s\ll1, v_r\ll1$, Eq.(\ref{metricrel}) will become that,
\begin{eqnarray}\label{}
&&ds^2=\frac{b\rho_0}{2c_s}\big[-(c_s^2-v^2_r)dt^2-2v_rdtdr+dr^2+r^2d\Omega^2\big],
\end{eqnarray}
which recovers the metric for non-relativistic fluid \cite{visser} with an overall factor $b/2$ difference.
It is obvious that this line element is conformal to the Schwarzschild geometry in the ingoing Painlev\'e-Gullstrand coordinates when $v_r=\sqrt{2M/r}$,
\begin{eqnarray}\label{}
&&ds^2=-f(r)dt^2+2\sqrt{\frac{2M}{r}}dtdr+dr^2+r^2d\Omega^2.
\end{eqnarray}
This conformal relation between these two metrics reveals a clear physical picture: when a perturbation (vibration) is applied to a non-relativistic fluid defined in Newtonian spacetime with a flow velocity of $\sqrt{2M/r}$, the vibration propagates in the fluid in the form of acoustic waves, and one will find that such propagation process occurs in a curved spacetime.

For the ingoing Painlev\'e-Gullstrand coordinates, the term $\sqrt{2M/r}$ is the local infall velocity of a free-falling particle that starts at rest at infinity and moves radially to the position with radius $r$ in Schwarzschild spacetime, which also corresponds to the radial dragging velocity of spacetime in Painlev\'e-Gullstrand coordinates.
Therefore we can set $v_r=\sqrt{2M\xi/r}$ as in Ref. \cite{visser} as the escape velocity of an observer maintaining a stationary position at radius $r$, where $\xi$ is a tuning parameter. From Eq. (\ref{rhom}), one can obtain $v_\mu v^\mu=m^2-1$ after re-scaling $m^2\rightarrow m^2/2c_s^2$, $v_\mu v^\mu\rightarrow v_\mu v^\mu/2c_s^2$. As addressed in \cite{ge2019}, we can work at the critical temperature of Gross-Pitaevskii theory such that $m^2$ vanishes, and $v_\mu v^\mu=-1$. So the time component of the velocity $v_t=\sqrt{f(r)+2M\xi f(r)^2/(1+\ell)r}$. It is easy to see that at infinite $r\rightarrow\infty$, $v_t\rightarrow1$, while $v_r\rightarrow0$.

Applying a coordinate transformation in Ref. \cite{ge2019,guo}
\begin{eqnarray}\label{}
&&dt\rightarrow dt-\frac{v_tv_r}{g_{tt}(c_s^2-v_rv^r)}dr,
\end{eqnarray}
then the line element (\ref{dtdr}) becomes
\begin{eqnarray}\label{metricsqr}
&&ds^2=c_s\sqrt{c_s^2-v_\mu v^\mu}\left[g_{tt}\frac{(c_s^2-v_rv^r)}{c_s^2-v_\mu v^\mu}dt^2+\frac{c_s^2}{(c_s^2-v_rv^r)}g_{rr}dr^2+r^2d\Omega^2\right].
\end{eqnarray}
After substituting metric functions of Eq. (\ref{metricbum}) into the above equation, the last line element of acoustic black hole is
\begin{eqnarray}\label{metric}
&&ds^2=G_{\mu\nu}dx^\mu dx^\nu=\sqrt{3}c_s^2\left[-F(r)dt^2+\frac{1+\ell}{F(r)}dr^2+r^2d\Omega^2\right],
\end{eqnarray}
with
\begin{eqnarray}\label{}
&& F(r)=f(r)\Big[1-\frac{f(r)}{(1+\ell)}v_r^2\Big]=f(r)\Big[1-\frac{2M\xi f(r)}{(1+\ell)r}\Big],
\end{eqnarray}
where $\xi$ is an artificially introduced dimensionless tuning parameter which guarantees that the acoustic horizon is outside the black hole event horizon. When $\ell\rightarrow0$, the acoustic black hole metric in Schwarzschild spacetime is recovered \cite{guo}. When $\xi\rightarrow0$ and $\sqrt{3}c_s^2=1$, the Schwarzschild-like black hole in Einstein-bumblebee theory is recovered \cite{casana}. The behaves of $F(r)$ with different constants $\xi$ and $ \ell$ is shown in Fig. 1.
\begin{figure}[ht]
\begin{center}
\includegraphics[width=7.0cm]{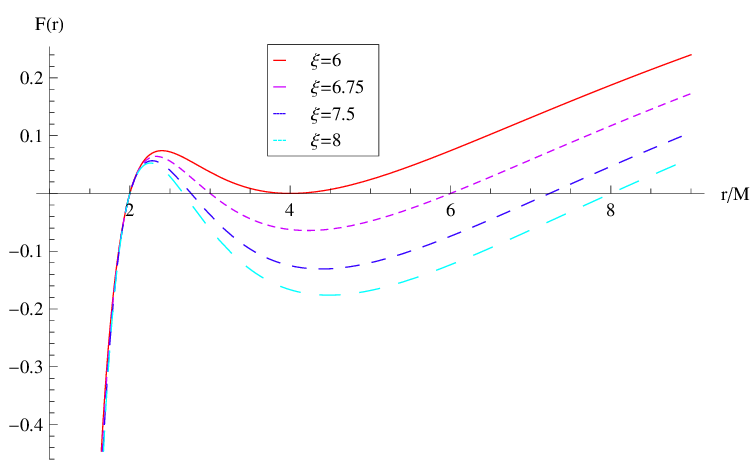}\;\;\;\;\includegraphics[width=7.0cm]{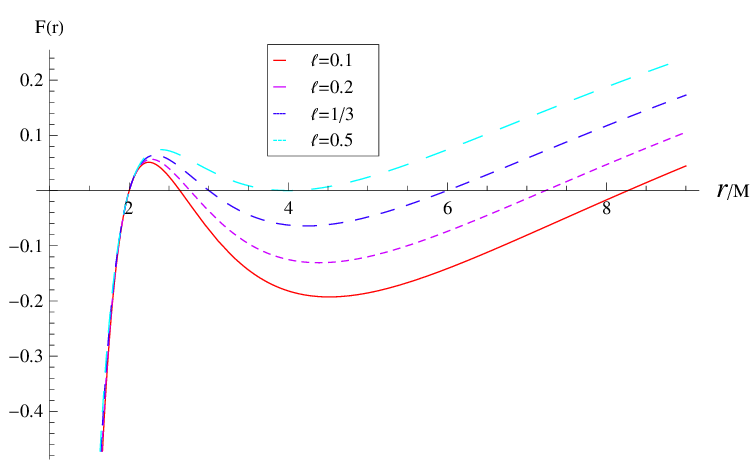}
\caption{The variation of the metric function $F(r)$ with different constant $\xi$ and $\ell$. In the right panel, $\ell$ is fixed to 0.5, $F(r)$ decreases with tuning parameter $\xi$; in the left panel, $\xi$ is fixed to 6, $F(r)$ increases with Lorentz violation coupling constant $\ell$. When $F(r)=0$, it gives the black hole horizon $r_h=2M$ and acoustic horizon $r_{ac}\geq4M,\; 2M<r_{ac-}\leq4M$. When $\xi=9(1+\ell)/2$, the acoustic horizon $r_{ac}=6M$.}\label{}
 \end{center}
 \end{figure}
Note that this metric $G_{\mu\nu}$ cannot solve Einstein's gravitational equation \cite{ge2019}.

 This acoustic Schwarzschild-like black hole has two kinds of horizons determined by $F(r)=0$: one is the gravitational(black hole) event horizon which is located at $r_h=2M$ (light behind it cannot escape out), the others are the acoustic event horizon $r_{ac}$ and inner acoustic horizon $r_{ac-}$ which are located at
\begin{eqnarray}\label{}
&&r_{ac}=\frac{M}{1+\ell}\big[\xi+\sqrt{\xi^2-4(1+\ell)\xi}\big],\quad r_{ac-}=\frac{M}{1+\ell}\big[\xi-\sqrt{\xi^2-4(1+\ell)\xi}\big].
\end{eqnarray}
To guarantee $r_{ac}$ or $r_{ac-}$ is real, the tuning parameter $\xi$ should be $\xi\geq4(1+\ell)$\footnote{When $0<\xi<4(1+\ell)$, the velocity $v_r$ of the background fluid is very small, then there is no acoustic horizon to exist, and in the region $r>2M$, the sound waves can always propagate out.}. It is easy to verify that $2M\leq r_{ac-}\leq4M$ and $r_{ac}\geq4M$, which are the same as those of the acoustic Schwarzschild black hole \cite{guo}. When $\xi=9(1+\ell)/2$, the acoustic event horizon $r_{ac}=6M$, which is the same as that of acoustic Schwarzschild black hole in the bath of the cosmological microwave background \cite{hartle,ge2019}. This location is also the same as that of the innermost stable circular orbit $r_{ms}=6M$ in Schwarzschild spacetime \cite{bardeen}.  The above properties are shown in Fig. 1. It shows us that the acoustic horizon $r_{ac}$ increases with the tuning parameter $\xi$, while decreases with the Lorentz violation parameter $\ell$. One can see that  the tuning parameter $\xi$ serves as the key control quantity to realize the special configuration where the acoustic horizon is located outside the optical event horizon of the real black hole.

Though the metric $G_{\mu\nu}$ in Eq. (\ref{metric}) isn't a solution of Einstein's equation, it shares all kinematics of General Relativity, such as Hawking temperature, Hawking radiation. Its Hawking temperature can be obtained by the equation that $T=F'(r_i)/4\pi\sqrt{1+\ell}$ \cite{ding2008}, where $r_i$ is a horizon position. For the Hawking temperature at the black hole horizon $r_h$, it is
\begin{eqnarray}\label{optic}
&&T_h=\frac{1}{8\pi\sqrt{1+\ell}M},
\end{eqnarray}
which is the same as that of Schwarzschild-like black hole \cite{gomes} and  entirely insensitive to the tuning parameter $\xi$.
For the Hawking temperature at the dump hole horizon $r_{ac}$, it is
\begin{eqnarray}\label{}
&&T_{ac}=\frac{M\xi(r_{ac}-2M)(r_{ac}-4M)}{2\pi\sqrt{1+\ell}^3r_{ac}^4},
\end{eqnarray}
which can recover the result of that in \cite{kanzi,yasir} when $\ell\rightarrow0$. The behaviors of $T_{ac}$ are shown in Fig. 2. It shows us that the tuning parameter $\xi$ tunes the fluid velocity and thereby tunes Hawking temperature of the phonon radiation.  As $\xi$ increases from a certain minimum value, the temperature rises rapidly from zero to a maximum and then decreases slowly with further increase of $\xi$. Whereas throughout the whole range in which $\ell$ grows from zero to a maximum value, $T_{ac}$  keeps decreasing until it finally vanishes. So the Hawking temperature $T_{ac}$ has a maximum value, which is  in accordance with the Maxwell-Boltzmann distribution \cite{kanzi,kaniadakis}.
\begin{figure}[ht]
\begin{center}
\includegraphics[width=7.0cm]{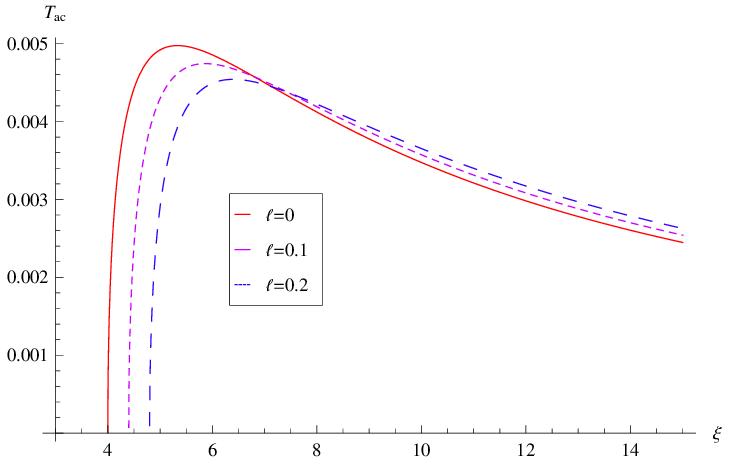}\;\;\;\;\includegraphics[width=7.0cm]{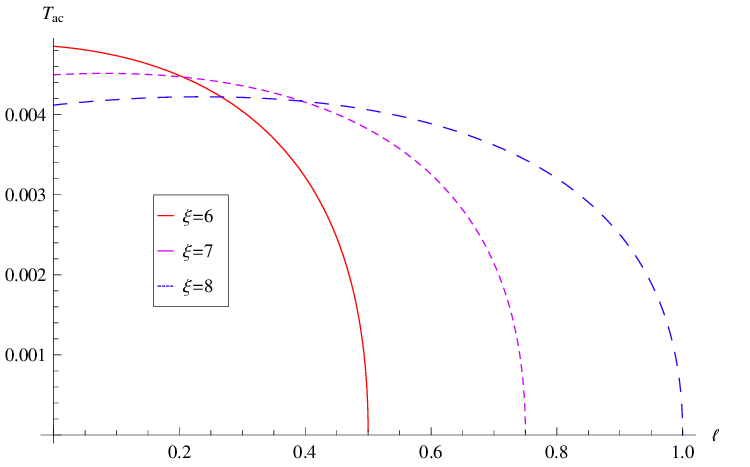}
\caption{The variation of Hawking temperature of the acoustic horizon $T_{ac}$ with different constant $\xi$ and $\ell$. In the right panel,  $T_{ac}$ is with tuning parameter $\xi$; in the left panel,  $T_{ac}$ is with Lorentz violation coupling constant $\ell$. Note that $\xi\geq 4(1+\ell)$ or $\ell\leq\xi/4-1$.}\label{}
 \end{center}
 \end{figure}
We can obtain this maximum value via the derivation of $T_{ac}$ with respect to $\xi$,
\begin{eqnarray}\label{}
&&\frac{\partial T_{ac}}{\partial \xi}=\frac{\sqrt{1+\ell}[\sqrt{\xi}-2\sqrt{\xi-4(1+\ell)}]}{8\pi M\xi^2\sqrt{\xi-4(1+\ell)}}.
\end{eqnarray}
When $\partial T_{ac}/\partial \xi=0$, it gives $\xi_m=16(1+\ell)/3$ and the fluid velocity is $v_{rm}=\sqrt{2M\xi_m/r}$, so the maximum value of the acoustic temperature $T_{acm}$ is
\begin{eqnarray}\label{}
&&T_{acm}=\frac{1}{64\pi \sqrt{1+\ell}M},
\end{eqnarray}
which is an eighth of $T_h$ of Eq. (\ref{optic}). So we learn that the acoustic temperature is smaller than the black hole temperature.
\section{Greybody factor}
In this section, we study the analogue Hawking radiation of this acoustic Schwarzschild-like black hole by considering it radiates massless scalar particles---phonons at the acoustic horizon. The physical picture can be described as following: the acoustic horizon splits the phonon modes into escaping and trapped branches whose vacua are related by a non-trivial Bogoliubov transformation; the exponential mode evolution then squeezes the vacuum, producing entangled phonon pairs of which one escapes as radiation and is thermal at $T_{ac}$. It shares the same kinematical core mechanism with the black hole radiation. But they are dynamically distinct in essentially every respect, the former is a phonon-field effect in an effective geometry induced by the medium and the latter is an effect of all quantum fields in a truly curved spacetime.

Although the escaping member of each entangled phonon pair is thermal at the acoustic Hawking temperature $T_{ac}$, the observed flux is modified by the greybody factor: the frequency-dependent tunnelling probability across the effective potential outside the horizon. We so evaluate this greybody factor and energy emission rate via WKB approximation. The minimally coupled massless scalar field equation is
\begin{eqnarray}\label{}
&&\frac{1}{\sqrt{-G}}\partial_\mu(\sqrt{-G}G^{\mu\nu}\partial_\nu\Phi)=0,
\end{eqnarray}
where $G_{\mu\nu}$ denotes the metric components of the acoustic Schwarzschild-like black holes (\ref{metric}). Setting $\Phi(t,r,\theta,\phi)=e^{-i\omega t}\mathcal{R}(r)Y_{lm}/r$ and defining the tortoise coordinate
\begin{eqnarray}\label{}
&&dr_*=\frac{\sqrt{1+\ell}}{F(r)}dr,
\end{eqnarray}
we find the radial equation for the scalar perturbation reads
\begin{eqnarray}\label{}
&&\frac{d^2\mathcal{R}(r)}{dr_*^2}+(\omega^2-V)\mathcal{R}(r)=0,\;\text{with}\;V=F(r)\left[\frac{l(l+1)}{r^2}
+\frac{F'(r)}{(1+\ell)r}\right],
\end{eqnarray}
with the boundary conditions that near the acoustic horizon $r_{ac}$, the purely incoming waves are allowed, while at infinity, the outgoing and incoming waves are allowed,
\begin{eqnarray}\label{}
&&\mathcal{R}(r)=Te^{-i\omega r_*},\;\;r_*\rightarrow-\infty;\qquad \mathcal{R}(r)=e^{-i\omega r_*}+Re^{i\omega r_*},\;\;r_*\rightarrow+\infty,
\end{eqnarray}
where $R$ and $T$ are the reflection and transmission coefficients satisfying $|T|^2+|R|^2=1.$
The radial domain of the following study is given by $r\in(r_{ac}, \infty)$. The effective potentials $V(r)$ for the fixed angular number $l=1$, different constants $\xi$ and $\ell$ are shown in Fig. 3. As approaching to the near acoustic horizon, the effective potential first shows a barrier and then quickly falls to zero at the horizon $r_{ac}$, meanwhile, the tortoise coordinate $r_*$ reaches minus infinity $r_*\rightarrow-\infty$.
\begin{figure}[ht]
\begin{center}
\includegraphics[width=7.0cm]{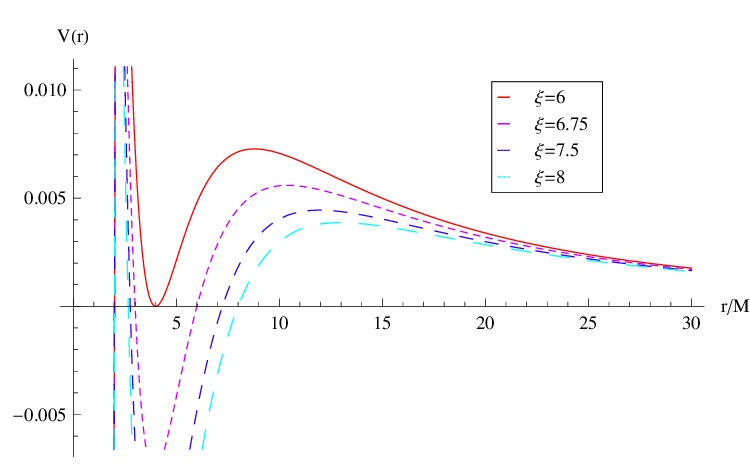}\;\;\;\;\includegraphics[width=7.0cm]{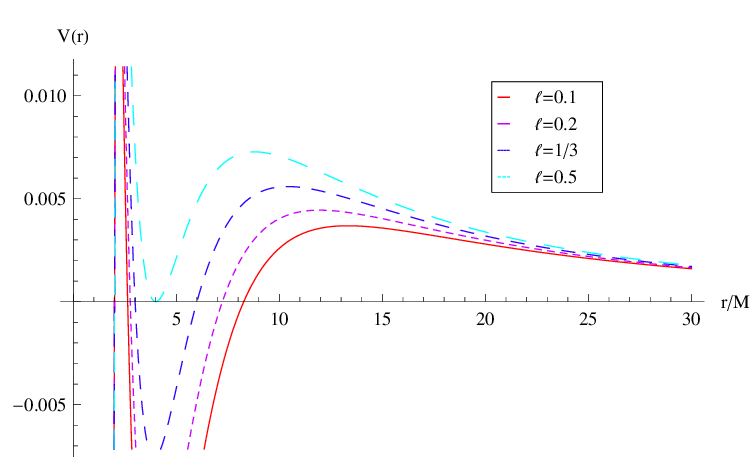}
\caption{The variation of effective potential $V(r)$ of the acoustic Schwarzschild-like black hole  with different constants $\xi$ and $\ell$. In the right panel,  it is the $V(r)$ with different tuning parameter $\xi$ and fixed $\ell=0.5$; in the left panel,  $V(r)$ with different Lorentz violation coupling constant $\ell$ and fixed $\xi=6$. Note that $\xi\geq 4(1+\ell)$ or $\ell\leq\xi/4-1$ and the angular number $l=1$ is fixed.}\label{}
 \end{center}
 \end{figure}
 The left plot of Fig. 3 shows the potential barrier is suppressed by larger tuning parameter $\xi$, which is similar to those of the  acoustic Schwarzschild  black hole \cite{guo} and will affect the Hawking radiation. The right plot of Fig. 3 the potential barrier increases with the Lorentz violating coupling constant $\ell$ which is similar to those of the modified acoustic black hole by considering the Abelian Higgs model with Lorentz symmetry breaking in flat Minkowski spacetime \cite{campos}.

The grey-body factor $|A_l|^2$ is then given by the transmission coefficient for each angular number as
\begin{eqnarray}\label{}
&&|A_l|^2=1-|R_l|^2=|T_l|^2\;\;\text{with} \;\;|T_l|^2=\frac{1}{1+e^{2\textrm{i}\pi K}},
\end{eqnarray}
where $K$ is determined by the equation
\begin{eqnarray}\label{}
&&K=\textrm{i}\frac{\omega^2-V_0}{\sqrt{-2V''_0}}-\sum_{i=2}^{6}\Lambda_i(K).
\end{eqnarray}
Here $V_0$ and $V''_0$ are the maximal value of the effective potential and its second derivative with respective to the tortoise coordinate at the maximum, respectively; and $\Lambda_i$ are the higher WKB corrections which are dependent on $K$ and up to 2$i$th-order of derivative of the potential at its maximum \cite{iyer,schutz,konoplya}.

Then we use the sixth-order WKB method \cite{konoplya2026} to calculate the grey-body factor $|A_l|^2$ of the Hawking radiation and show them in Fig. 4 only for the angular number $l=0$. This is because that the mode with angular number $l=0$ dominates the Hawking radiation, while the contribution of modes with higher $l$ is very small and negligible.
\begin{figure}[ht]
\begin{center}
\includegraphics[width=7.0cm]{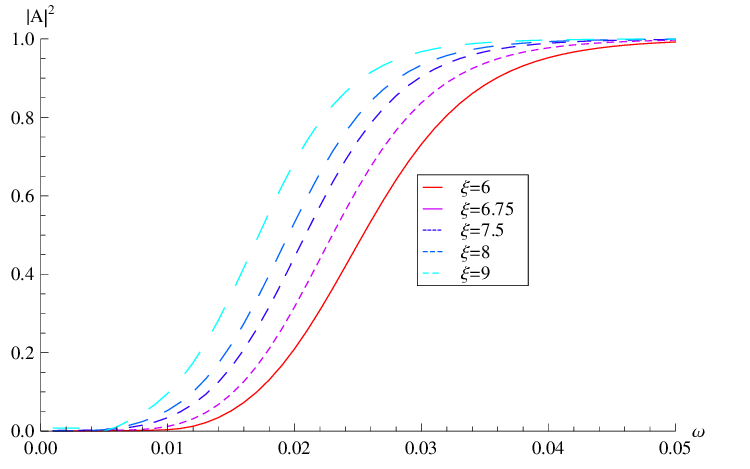}\;\;\;\;\includegraphics[width=7.0cm]{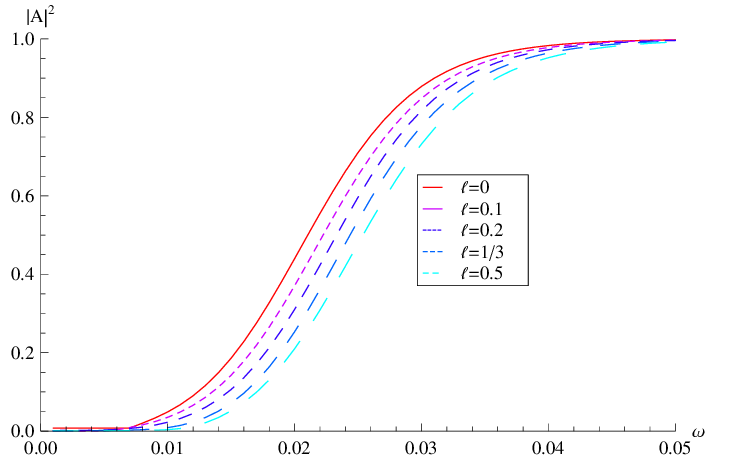}
\caption{The variation of grey-body  factor $|A|^2$ of the acoustic Schwarzschild-like black hole with different constant $\xi$ and $\ell$. In the right panel,  it is the $|A|^2$  with different tuning parameter $\xi$ and fixed $\ell=0.5$; in the left panel,  it is $|A|^2$ with the different Lorentz violation coupling constant $\ell$ and fixed $\xi=6$. Note that $\xi\geq 4(1+\ell)$ or $\ell\leq\xi/4-1$ and the angular number $l=0$ is fixed.}\label{}
 \end{center}
 \end{figure}
Fig. 4 shows that the larger frequency corresponds to the higher grey-body factor, because the particles with larger energy are more likely to penetrate the potential barrier. Its left panel shows that when one increases the tuning parameter $\xi$, the corresponding grey-body also increases, which are similar to those of the acoustic Schwarzschild black hole \cite{guo}. On the contrary, its right panel shows that when one increases the Lorentz violating parameter $\ell$, the corresponding grey-body decreases.

Nextly, we can study the energy emission rate of the Hawking radiation via the given grey-body factor by the following connection \cite{hawking},
\begin{eqnarray}\label{}
&&\frac{dE}{dt}=\sum_lN_l|A_l|^2\frac{\omega}{e^{\omega/T_{ac}}-1}\frac{d\omega}{2\pi},
\end{eqnarray}
where $N_l$ are the multiplicities satisfying $N_l=2l+1$ for the scalar field. Fig. 5 shows the energy emission rate of the Hawking radiation for the angular number $l=0$ mode with the different constants $\xi$ and $\ell$.
\begin{figure}[ht]
\begin{center}
\includegraphics[width=7.0cm]{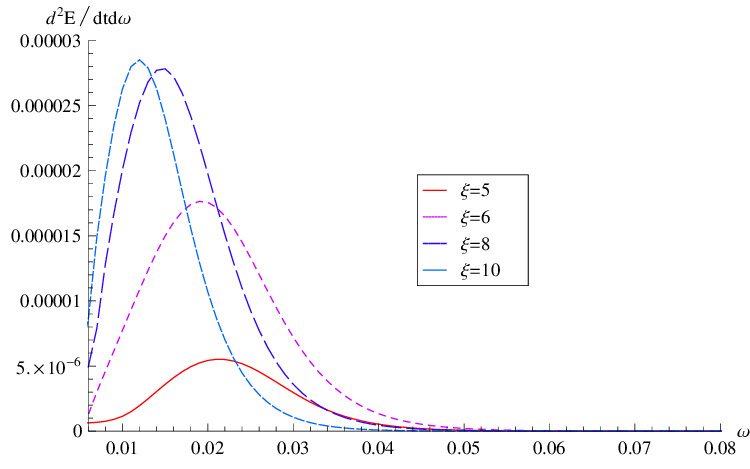}\;\;\;\;\includegraphics[width=7.0cm]{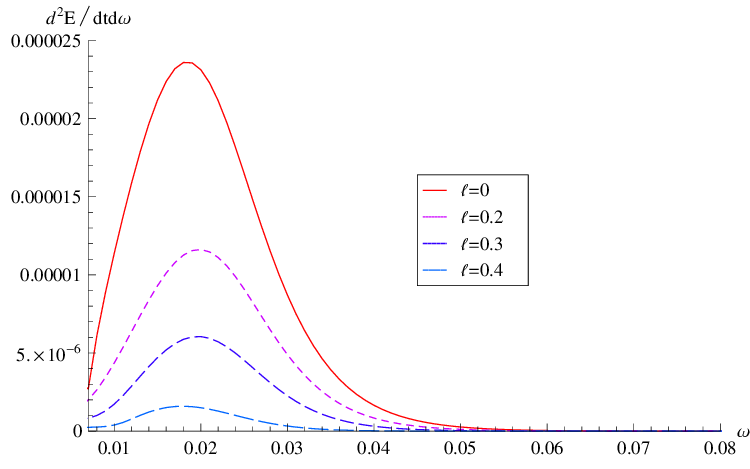}
\caption{The variation of energy radiation rate $d^2E/dtd\omega$ of the acoustic Schwarzschild-like black hole with different constant $\xi$ and $\ell$. In the right panel,  it is the $d^2E/dtd\omega$  with different tuning parameter $\xi$ and fixed $\ell=0.1$; in the left panel,  it is $d^2E/dtd\omega$ with the different Lorentz violation coupling constant $\ell$ and fixed $\xi=6$. Note that $\xi\geq 4(1+\ell)$ or $\ell\leq\xi/4-1$ and the angular number $l=0$ is fixed.}\label{}
 \end{center}
 \end{figure}
 Its left panel shows that the energy emission rate is larger for the larger tuning parameter $\xi$, which is similar to those of acoustic Schwarzschild black hole \cite{guo}; while the right panel shows that the energy emission rate is smaller for the larger Lorentz violating constant $\ell$. These energy emission rates are similar to Planck's black body spectrum given by Planck formula \cite{bernstein} that all of which have a maximum at a frequency $\omega_{m}$. These peak frequencies $\omega_m$ all decrease when the Hawking temperature $T_{ac}$ decreasing.

\section{Summary and discussion}

Traditional artificial acoustic black holes are constructed exclusively in a flat spacetime background and can only simulate a single acoustic horizon. In contrast, an acoustic black hole embedded in a curved spacetime background naturally possesses two independent horizons, namely the gravitational horizon and the acoustic horizon. The interaction between these two horizons can reproduce the transonic dynamical processes of the accretion flow around a real astrophysical black hole. So in this paper, we derive an acoustic Schwarzschild-like black hole in Einstein-bumblebee theory by using Gross-Pitaevskii theory in the background of Schwarzschild-like black hole spacetime.

This black hole has two kinds of event horizons: one is the gravitational horizon $r_h=2M$(black hole horizon, in which the light and sound cannot escape out), the others are the internal and external acoustic  horizons $2M<r_{ac-}\leq4M,\;r_{ac}\geq4M$(dumb hole horizon, in which only the sound cannot escape out) which depend on the tuning parameter $\xi$ and Lorentz violation coupling constant $\ell$. When the tuning parameter $\xi=4(1+\ell)$, the both acoustic  horizons overlap to extreme case that $r_{ac-}=r_{ac}=4M$. When $0<\xi<4(1+\ell)$, there is no acoustic horizon to exist. When we set $\xi=9(1+\ell)/2$, the horizon $r_{ac}=6M$, which is the same as that of the acoustic Schwarzschild black hole in the bath of the cosmological microwave, or that of the innermost stable circular orbit in Schwarzschild black hole spacetime. The corresponding Hawking temperature $T_h$ depends only on $\ell$, however $T_{ac}$ depends on $\ell$ and $\xi$. The dumb hole temperature $T_{ac}$ has a maximum $T_{acm}=T_h/8$ when the tuning parameter is set to be $\xi=16(1+\ell)/3$.

This acoustic black hole shares all kinematics of General Relativity. So when considering the scalar particles radiating from the dumb hole, we find that in low frequency $\omega<0.08$, the grey-body factor increases with the tuning parameter $\xi$, which is similar to those of the acoustic Schwarzschild black hole; but it decreases with Lorentz violation parameter $\ell$. In the high frequency, it approaches to $1$ for all $\xi$ and $\ell$. Its energy emission rate is larger for the larger tuning parameter $\xi$, which is similar to those of the acoustic Schwarzschild black hole; however it decreases with Lorentz violation parameter $\ell$. And this energy emission rate is similar to the Planck's black spectrum where there has a maximum at a given peak frequency $\omega_m$. This peak frequency decreases with $\xi$ or $\ell$.

The above parameter $(\xi,\ell)$-dependent law is the core physical characteristic of the Lorentz-violating acoustic Schwarzschild-like  black hole system. The physical essences of the two parameters are completely different. $\xi$ is the hydrodynamic parameter,  it describes the radial velocity of the accretion flow. An increase in $\xi$ means that the fluid velocity rises more rapidly, and the acoustic horizon will move outward away from the gravitational horizon; the height of the effective scattering potential barrier is reduced, as a result, the grey-body factor increases, more phonons can pass through the barrier and radiate outward.
On the other hand, $\ell$ is the gravitational correction parameter: it corresponds to the strength of Lorentz symmetry breaking in the Einstein-bumblebee model. An increase in $\ell$ will directly modify the geometric structure of the background spacetime, enhancing the equivalent gravitational drag effect of the radial metric term. Then the height of the effective potential barrier rises, consequently, the greybody factor decreases, phonons find it harder to penetrate the barrier and escape outward.

This opposite dependence relationship is the most experimentally valuable feature of this system: in laboratory acoustic black hole experiments, you can independently adjust the fluid flow parameters $\xi$ and the background equivalent spacetime parameters $\ell$, to completely distinguish the contributions of hydrodynamic effects and Lorentz-violating gravitational correction effects to Hawking radiation. This solves the problem that the two effects are coupled and cannot be measured separately in traditional acoustic black holes with a flat spacetime background.

This relationship can be also directly benchmarked against observations of real astrophysical black holes: if future observations via the EHT black hole shadow and LIGO gravitational wave detection find that the black hole radiation rate is lower than the prediction of General Relativity, it can correspond to a positive correction of the Lorentz violation parameter $\ell$; if the observed radiation rate is higher than the General Relativity prediction, it corresponds to the variation of the $\xi$ parameter in the accretion flow velocity profile. The observational signatures of the two are completely opposite without degeneracy, providing a clear discriminant basis for the experimental/observational test of Lorentz symmetry breaking.

\begin{acknowledgments}  Zhou's work was supported by the National Natural Science Foundation of China(NNSFC) under  Grant No. 12375047; Shi's work was supported by NNSFC under  Grant No. 12305063.
\end{acknowledgments}

\appendix\section{Canonical acoustic black hole with Lorentz violation}

In this appendix, we construct a canonical acoustic black hole in the background of Schwarzschild-like black hole spacetime. It is the most natural and minimal assumptions that a spherically symmetric flow is incompressible, then the density $\rho$ is position independent and so is the pressure and the speed of sound $c_s$ as well. Since $\rho$ is position independent, the continuity equation implies $v_r\propto1/r^2$. So we define the flow speed as
\begin{eqnarray}\label{vrzeta}
&&v_r=c_s\frac{\zeta}{r^2},
\end{eqnarray}
where $\zeta$ is a tuning parameter which guarantees that the acoustic horizon is outside the the optical horizon. After substituting it into Eq. (\ref{metricsqr}), we can obtain the metric function of the canonical acoustic black hole
\begin{eqnarray}\label{}
&&F(r)=f(r)\left[1-\frac{\zeta f(r)}{(1+\ell)r^4}\right].
\end{eqnarray}
If $f(r)\rightarrow1$ and $\ell\rightarrow0$, then it can be reduced to the result of Ref. \cite{visser}.
The equation $F(r)=0$ gives the optical horizon $r_h=2M$ and the acoustic horizon $r_{ac}$. To guarantee $r_{ac}>2M$, the parameter $\zeta$ should  satisfies the condition $\zeta\geq\zeta_c,$ where $\zeta_c=5(1+\ell)\big(5M/2\big)^4$. It is easy to verify that the internal acoustic horizon $2M<r_{ac-}<5M/2$, and the external horizon $r_{ac}>5M/2$. When $\zeta=\zeta_c$, the internal and external acoustic horizons coincide at $r_{ac}=5M/2$.

\end{document}